\documentclass[fleqn,usenatbib]{mnras}

\usepackage{newtxtext,newtxmath}

\usepackage[T1]{fontenc}

\DeclareRobustCommand{\VAN}[3]{#2}
\let\VANthebibliography\thebibliography
\def\thebibliography{\DeclareRobustCommand{\VAN}[3]{##3}\VANthebibliography}

\usepackage{graphicx}	
\usepackage{amsmath}	
\usepackage{siunitx}
\usepackage{ulem}

\title[Spatial variation in local LyC leaker]{Understanding the spatial variation of \ion{Mg}{II} and ionizing photon escape in a local LyC leaker}

\author[T. Seive et al.]{
Thomas Seive,$^{1}$\thanks{E-mail: tseive22@gmail.com}
John Chisholm,$^{1}$\thanks{E-mail: chisholm@austin.utexas.edu}
Floriane Leclercq,$^{1}$
Gregory Zeimann,$^{1}$
\\
$^{1}$Department of Astronomy, University of Texas, Austin, TX 78712, USA\\
}

\date{Accepted XXX. Received YYY; in original form ZZZ}

\pubyear{2022}

\begin{document}
\label{firstpage}
\pagerange{\pageref{firstpage}--\pageref{lastpage}}
\maketitle

\begin{abstract}
Ionizing photons must have escaped from high-redshift galaxies, but the neutral high-redshift intergalactic medium makes it unlikely to directly detect these photons during the Epoch of Reionization. Indirect methods of studying ionizing photon escape fractions present a way to infer how the first galaxies may have reionized the universe. Here, we use HET/LRS2 observations of J0919+4906, a confirmed z$\approx$0.4 emitter of ionizing photons to achieve spatially resolved (12.5 kpc in diameter) spectroscopy of \ion{Mg}{II}$\lambda2796$, \ion{Mg}{II}$\lambda2803$, [\ion{O}{II}]$\lambda\lambda3727,3729$ , [\ion{Ne}{III}]$\lambda3869$, H$\gamma$, [\ion{O}{III}]$\lambda4363$, H$\beta$, [\ion{O}{III}]$\lambda4959$, [\ion{O}{III}]$\lambda5007$, and H$\alpha$. From these data we measure \ion{Mg}{II} emission, which is a promising indirect tracer of ionizing photons, along with nebular ionization and dust attenuation in multiple spatially-resolved apertures. We find that J0919+4906 has significant spatial variation in its \ion{Mg}{II} escape and thus ionizing photon escape fraction. Combining our observations with photoionization models, we find that the regions with the largest relative \ion{Mg}{II} emission and \ion{Mg}{II} escape fractions have the highest ionization and lowest dust attenuation. Some regions have an escape fraction that matches that required by models to reionize the early universe, while other regions do not. We observe a factor of 36 spatial variation in the inferred LyC escape fraction, which is similar to recently observed statistical samples of indirect tracers of ionizing photon escape fractions. These observations suggest that spatial variations in neutral gas properties lead to large variations in the measured LyC escape fractions. Our results suggest that single sightline observations may not trace the volume-averaged escape fraction of ionizing photons.


\end{abstract}




\section{Introduction}
\label{sec:intro}
Between redshifts of 5-10, the intergalactic medium (IGM) in the early universe rapidly went through a phase transition from neutral to ionized \citep{Becker_2001,Fan_2006,Banados_2017,Becker_2021}. This ionization was brought about by Lyman Continuum (LyC) photons which have $\lambda < 912$ \AA, escaping from galaxies. Understanding the source of these photons will reveal the timing, duration, and overall evolution of large scale structure in the universe \citep{Robertson_2013,Madau_2014,Robertson_2015}. Furthermore, the process of finding the sources of ionizing photons can inform us about the quenching of dwarf galaxies in the early universe \citep{Bullock_2000} and explain the IGM temperature \citep{Miralda_1994}. Thus, understanding the LyC escape is crucial for establishing the observable universe.

Initially, the debate about the sources of reionization centered around whether active galactic nuclei (AGN) or massive stars were most responsible for providing ionizing photons \citep{Ouchi_2009a,Faucher_2009,Robertson_2013,Madau_2015}. AGN generate many ionizing photons and are concentrated in the depths of gravitational potentials. However, current observations find that there are too few AGNs to reionize the universe \citep{Hopkins_2008,Onoue_2017,Ricci_2016,Matsuoka_2018,Shen_2020}. On the other hand, star forming galaxies are readily observed at the required redshifts and are broadly distributed spatially. However, in order to be star-forming, they must have large amounts of cold gas. This cold, neutral gas will efficiently absorb ionizing photons, nearly eliminating the photons that escape a typical star-forming galaxy. This reduction in escape fraction from star-forming galaxies has shifted the current debate to whether widely distributed low-mass galaxies or heavily clustered bright massive galaxies were the sources of ionizing photons \citep{Finkelstein_2019,Naidu_2020,matthee_2021}. The difference in concentration between the various types of ionizing photon sources will affect the morphology of reionization. Infrequent, concentrated sources (e.g. AGNs and massive galaxies) will yield a much patchier evolution of reionization when compared to the less bright and more evenly distributed sources (e.g. low-mass galaxies; \citealt{Rosdahl_2018}).

The number density of ionizing photons that escape a source, known as the ionizing emissivity (J\textsubscript{ion} [photon s\textsuperscript{-1} Mpc\textsuperscript{-3}]), requires measurements of the density of sources and the production of ionizing photons of each source. J\textsubscript{ion} can be numerically represented as 
\begin{equation}
    J\textsubscript{ion} = f\textsubscript{esc}(\textnormal{LyC}) \xi\textsubscript{ion} \rho\textsubscript{UV}
\end{equation}
where f\textsubscript{esc}(LyC) is the volume-averaged fraction of ionizing photons that escape from all sides of a galaxy, $\xi$\textsubscript{ion} is the intrinsic production of ionizing photons per far ultraviolet (FUV) luminosity for each source, and $\rho$\textsubscript{UV} is the co-moving UV luminosity density, which has been derived from the luminosity function. Observing all of these parameters for star forming galaxies in the Epoch of Reionizion (EoR) would answer how the universe became ionized. Unfortunately, all of these parameters have their own challenges associated with measuring them, with escape fractions currently being the most uncertain of them.


Confirming whether star forming galaxies emitted a sufficient number of ionizing photons to reionize the Universe requires observations of the total volume-averaged LyC escape fraction of a galaxy. However, observing total volume-averaged LyC escape fractions at redshifts between z $\approx 4$ and the EoR would be difficult because of intervening neutral gas. Also, the fact that observations can only probe the portion of the galaxy facing the observer makes a total volume-averaged LyC escape fraction impossible to measure for a given galaxy.  Recently, there has been tremendous success with directly measuring the escape of ionizing photons through single sightlines at low-redshift, with measurements of escape fractions between 0-70\% (e.g. \citealt{Grimes_2009,Vanzella_2010,Leitet_2011,Borthakur_2014,Izotov_2016a,Izotov_2016b,Leitherer_2016,Shapley_2016,Vanzella_2016,Izotov_2018a,Izotov_2018b,Naidu_2018,Steidel_2018,Fletcher_2019,Rivera_2019,Wang_2019,Izotov_2021, Flury_2022, Flury_2022b}). Concurrently, indirect tracers of LyC escape have proven to be a promising route to penetrate the barriers at the EoR. Thus far, these methods have included Ly$\alpha$ emission properties \citep{Verhamme_2015,Rivera_2017,Izotov_2018b,Gazagnes_2021,Izotov_2021,Flury_2022b}, ISM absorption properties \citep{Reddy_2016,Gazagnes_2018,Chisholm_2018,Steidel_2018,Saldana_Lopez_2022}, resonant emission lines \citep{Henry_2018,Chisholm_2020,Witstok_2021,Schaerer_2022, Izotov_jul_2022, Xu_2022}, and optical emission line ratios \citep{Nakajima_2014,Oey_2014,Wang_2019,Flury_2022b}. However, from \citealt{Flury_2022b}, many of these indirect indicators have substantial scatter, making it challenging to indirectly infer the LyC escape fraction from single observations. The large scatter in escape fractions has been explored in simulations and connected to large spatial scatter in f\textsubscript{esc}(LyC) \citep{Trebitsch_2017,Rosdahl_2018,Mauerhofer_2021}. Of all of these tracers, the resonant line \ion{Mg}{II} has shown great promise as a tracer of LyC escape \citep{Henry_2018,Chisholm_2020}. Due to being a resonant line, emission from the \ion{Mg}{II} doublet has been found to trace neutral gas at column densities less than $10^{17}$ cm\textsuperscript{-2} \citep{Chisholm_2020}. This makes it capable of determining a neutral gas column density low enough to transmit ionizing photons. Along with this, the ionization energy of \ion{Mg}{II} overlaps with that of neutral hydrogen (15 eV and 13.6 eV, respectively). For all these reasons, \ion{Mg}{II} emission has been suggested to be an ideal indirect indicator of the escape of ionizing photons.

All of these methods have had major work to either directly measure escape fractions or use observations to stringently test indirect methods. However, direct or indirect, all the previously mentioned measurements have all been along single sightlines centered on the brightest regions of galaxies. These single sightlines are one out of many ways to observe the galaxy. As mentioned above, an understanding of reionization requires the total amount of ionizing photons that escape from all sides of a galaxy and reach the intergalactic medium, also known as a volume averaged escape fraction. Without a volume averaged escape fraction, it is challenging to extrapolate single sightline LyC measurements to observations of the EoR. From this challenge, a series of questions naturally arise: (a) How can we interpret a single sightline LyC observation? (b) Does this single sightline observation relate to the volume averaged escape fraction? (c) Does spatial variation obscure the picture created by single sightline observations?


Here we aim to answer these questions and test the spatial variation of the neutral gas opacity in a previously confirmed z $\approx0.4$, LyC emitting galaxy \citep{Izotov_2021}. We use spatially resolved Integrated Field Unit (IFU) observations from the Low Resolution Spectrograph 2 (LRS2) instrument \citep{Chonis_2014} on the Hobby-Eberly Telescope (HET), to determine \ion{Mg}{II} along with other emission lines and test if spatial variation in \ion{Mg}{II} flux, dust, and ionization in a target can lead to substantial variation in sightline to sightline escape fractions. In \autoref{sec:observationsandreductions} we describe the LRS2 observations and data reduction. Section \ref{sec:methods} describes our process for extracting emission line parameters and draws similarities and differences to other observations of this galaxy in the literature. Our observations of the dust, ionization, and \ion{Mg}{II} emission are described in \autoref{sec:results}. We conclude in \autoref{sec:discussion} by exploring the implications of our measurements for single sightline observations. Throughout the paper, all distances are in physical units, not comoving. We assume a flat CDM cosmology with $\Omega_{m}$ = 0.315 and $H_0$ = 67.4 km s\textsuperscript{-1} Mpc\textsuperscript{-1} \citep{Planck_2020}; in this framework, a 1~${\rm arcsec}$ angular separation corresponds to 5 kpc proper at the redshift of the galaxy.

\section{Observations and reductions}
\label{sec:observationsandreductions}

\subsection{Observations}
\label{sub:observations}
We observed J0919+4906 (hereafter J0919, RA: 09:19:55.78, Dec: +49:06:08.75) over 3 nights (January 8th 2021, January 9th 2021, and March 3rd 2021) with four total exposures using the LRS2 spectrograph \citep{Chonis_2014,Chonis_2016} on the Hobby-Eberly Telescope. This object is of interest because it was one of the recently discovered Lyman Continuum (LyC) emitters from \citet{Izotov_2021}. J0919 has a stellar mass of $10^{7.51}$M\textsubscript{\(\odot\)}, a star formation rate of 8.4 M\textsubscript{\(\odot\)}yr\textsuperscript{-1}, and a redshift of 0.40512 \citep{Izotov_2021}. Three of the exposures had an exposure time of 1800 seconds using the LRS2-B configuration. The fourth exposure had an exposure time of 500 seconds with the LRS2-R configuration. The maximum seeing was a FWHM of $\approx$2.6~${\rm arcsec}$ and the final data cube had a spatial scale of 0.25~${\rm arcsec}$ by 0.25~${\rm arcsec}$ spaxels. The spectral resolution of the observation depends on the LRS2 spectrograph arm (see \autoref{sub:datareduction} for a description of the arms; UV: 1.63\AA\ , Orange: 4.44\AA\ , Red: 3.03\AA\ , Farred: 3.78\AA ).

This object has been observed with the SDSS in the optical \citep{SDSS_2019} and HST/COS in the FUV \citep{Izotov_2021}. These observations supplement our work. The SDSS observation provides values to compare against the LRS2 observations (\autoref{sub:sdss check}). From our central aperture (\autoref{fig:aperture_figure}), we measure a Signal to Noise Ratio (SNR) of 13.5 in \ion{Mg}{II}$\lambda2796$, which is one of our weakest emission lines. This SNR is $\sim$1.2 times higher than the SNR of 11 from the SDSS for the same line. The HST/COS observations provide a direct measurement of the LyC escape fraction of 16\% \citep{Izotov_2021}. The LyC escape fraction is a value we attempt to indirectly measure in this work (\autoref{sub:IDE}).

\subsection{Data reduction}
\label{sub:datareduction}
The HET observations reported here were obtained with the LRS2 spectrograph. LRS2 comprises two spectrographs separated by 100 arcseconds on sky: LRS2-B (with wavelength coverage of 3650\AA\ -- 6950\AA) and LRS2-R (with wavelength coverage of 6450\AA\ -- 10500\AA).  Each spectrograph has 280 fibers, each with a diameter of 0.59~${\rm arcsec}$, covering  6~${\rm arcsec}$ $\times~$12~${\rm arcsec}$ with nearly unity fill factor \citep{Chonis_2014}. We used the HET LRS2 pipeline, Panacea\footnote{https://github.com/grzeimann/Panacea}, to perform the initial reductions including: fiber extraction, wavelength calibration, astrometry, and flux calibration. There are two channels for each spectrograph: UV and Orange for LRS2-B and Red and Farred for LRS2-R. On each exposure, we combined fiber spectra from the two channels into a single data cube accounting for differential atmospheric refraction. We then identified the target galaxy in each observation and rectified the data cubes to a common sky coordinate grid with the target at the center.

For sky subtraction, we take the biweight spectrum of all spaxels >4~${\rm arcsec}$ from the target to minimize self-subtraction. This 4~${\rm arcsec}$ aperture was chosen by looking at the strongest emission lines and determining their spatial extents by eye. To further test our data reduction, we fit 2D circular Gaussians to the [\ion{O}{III}]$\lambda5007$ and [\ion{O}{II}]$\lambda\lambda3727,3729$ spatial maps and find that more than 99\% of the flux for both lines was within 4~${\rm arcsec}$. From this test we find that the 4~$\rm arcsec$ mask for the sky subtraction is well chosen. At each wavelength in the new cube we perform an additional residual sky subtraction using a 2.5~${\rm arcsec}$ Gaussian smoothing kernel and a 2.5~${\rm arcsec}$ mask of the target to model the coherent sky-line residuals that arise from the spatially varying spectral resolution caused by differences in instrumental performance from fiber to fiber \citep{Chonis_2016}.

To normalize each cube, we measured H$\beta$ in both the LRS2-B and LRS2-R IFUs at the observed wavelength of $\approx$6831\AA. After normalization we stacked the individual cubes together using a variance weighted mean. In \autoref{sub:sdss check}, we check the flux calibration by comparing the line ratios of the SDSS and LRS2 spectra and find offsets ranging from $0\sigma$ to $3\sigma$ between the two datasets. We correct the wavelengths to the heliocentric frame.

Our final test was to determine the impact of wavelength dependent calibrations on the trends in \autoref{sec:results}. To accomplish this, we compared the [\ion{Ne}{III}]/[\ion{O}{II}] ratio to the [\ion{O}{III}]/[\ion{O}{II}] ratio. Both of these ratios trace ionization but are at very different wavelength separations. This test demonstrates that even at these different wavelength separations both ratios display similar trends. Therefore, it is not wavelength dependent calibrations that drive the trends in \autoref{sec:results}.


\section{Emission line parameter estimation}
\label{sec:methods}
Here we describe the techniques used to derive emission line properties from the observations described in \autoref{sec:observationsandreductions}. We first defined apertures for the spectral extraction (\autoref{sub:aperture}), removed the continuum from the spectra (\autoref{sub:cont_fit}), fit the emission lines (\autoref{sub:emission}), and corrected for dust attenuation (\autoref{sub:dust}). Our analysis was primarily done using the \textsc{specutils}, \textsc{spectralcube} \citep{astropy:2018}, and \textsc{lmfit} \citep{lmfit} python packages.

\subsection{Apertures}
\label{sub:aperture}
A primary goal of this work is to test the impact of geometry and spatial distribution on the resonant \ion{Mg}{II} emission. To do this, we extract the spectral information from spatially distinct apertures within the LRS2 data cube that are separated by more than the convolved seeing of the observations. Doing this leaves us with 5 spaxel radius (1.25~${\rm arcsec}$ radius, 12.5 kpc diameter) apertures, as dictated by the seeing of the observations (\autoref{sub:observations}). Given this aperture size, we optimized the number of apertures while maximizing the delivered SNR by extracting our signal from four spatially distinct regions. These regions covered the extent of J0919 without overlapping in the center, as seen in \autoref{fig:aperture_figure}.

While the center-most aperture (radius of 1~${\rm arcsec}$) does overlap with the other apertures, it allows for the LRS2 data to be compared to the SDSS optical spectra from the BOSS spectrograph. With this same purpose in mind, we created a COS analogue aperture (radius of 1.25~${\rm arcsec}$) centered on the galaxy to compare the LRS2 LyC escape fraction to \citealt{Izotov_2021}. The large aperture, which has a radius of 3.125~${\rm arcsec}$ and is referred to as the "Integrated aperture", maximizes the SNR of our observations and also provides the integrated spectrum of the entire galaxy. We extracted the spectra by summing the flux in every spaxel of each aperture. \autoref{fig:overlay_figure} shows the extracted spectra for various lines for each individual aperture. 

\subsection{Continuum fit}
\label{sub:cont_fit}
To ensure the measured emission line properties did not contain contributions from the stellar and nebular continua, we first had to remove the continuum. For our continuum fitting procedure, we used the \textsc{fit\_continuum} function from the \textsc{specutils} package. We modified the default Chebyshev model to be 1st order instead of 3rd order to better match the observed continuum shape. In order to fit the continuum, we visually picked a region of approximately 100~\AA\ on either side of the emission line that was in close proximity to the line but did not contain any absorption/emission features. This procedure was applied to the spectra extracted in the different apertures. The residual emission flux was then obtained by subtracting the resulting continuum models.

J0919 has an extreme H$\beta$ equivalent width (EW) of 435~\AA\ \citep{Izotov_2021}, indicating that its stellar population is very young. According to stellar population models, such young populations have up to $\sim$2~\AA\ H$\beta$ EW \citep{Delgado_1999}, which represents less than 0.5\% of J0919's total H$\beta$ EW. Further, with the continuum not being significantly detected underneath the Balmer lines (see \autoref{fig:fitexamples_figure} bottom row, third panel), we conclude that there is little contribution from stellar absorption. This is confirmed when we explored the requirements of stellar absorption on the measured Balmer emission line fluxes in J0919 and found that only the Integrated and Western apertures needed a stellar absorption correction (see \autoref{sub:dust}). Additionally, in the \ion{Mg}{II} region, there could be photospheric absorption from the stars \citep{Martin_2009,Henry_2018}. This typically occurs for A and F type stars which are much less massive than the stars that dominate the continua in J0919 \citep{Snow_1994}.

\subsection{Emission line fit}
\label{sub:emission}
Our method to measure the emission line parameters consisted of using \textsc{lmfit} \citep{lmfit} and a bootstrap Monte Carlo method. More precisely, we use the \textsc{minimize()} function and a Gaussian model with parameters of line center, line width, and amplitude to achieve all of our fits. The lower limit for the line width of our fits was based on the spectral resolution of each LRS2 arm (see \autoref{sub:observations} for the limits). While our model returns many parameters, this work focuses on the integrated fluxes. Subsequent work will study the kinematic information of the emission lines. 

For our bootstrap Monte Carlo method,  we first estimated the noise level by calculating the standard deviation of the continuum-subtracted data (see \autoref{sub:cont_fit}) in two 80-100 pixel-wide spectral windows directly adjacent to each individual emission line. With the \textsc{numpy.random.normal()} function \citep{numpy}, we generated 1000 realizations of the extracted spectrum where each flux density is randomly drawn from a normal distribution centered on the original flux density value with a standard deviation given by the estimated noise value calculated above. We then fit a Gaussian to each of the 1000 modified spectra. Our initial values for the models came from the \textsc{specutils} \textsc{find\_lines\_threshold} function. 

\begin{figure*}
	\includegraphics[width=\textwidth]{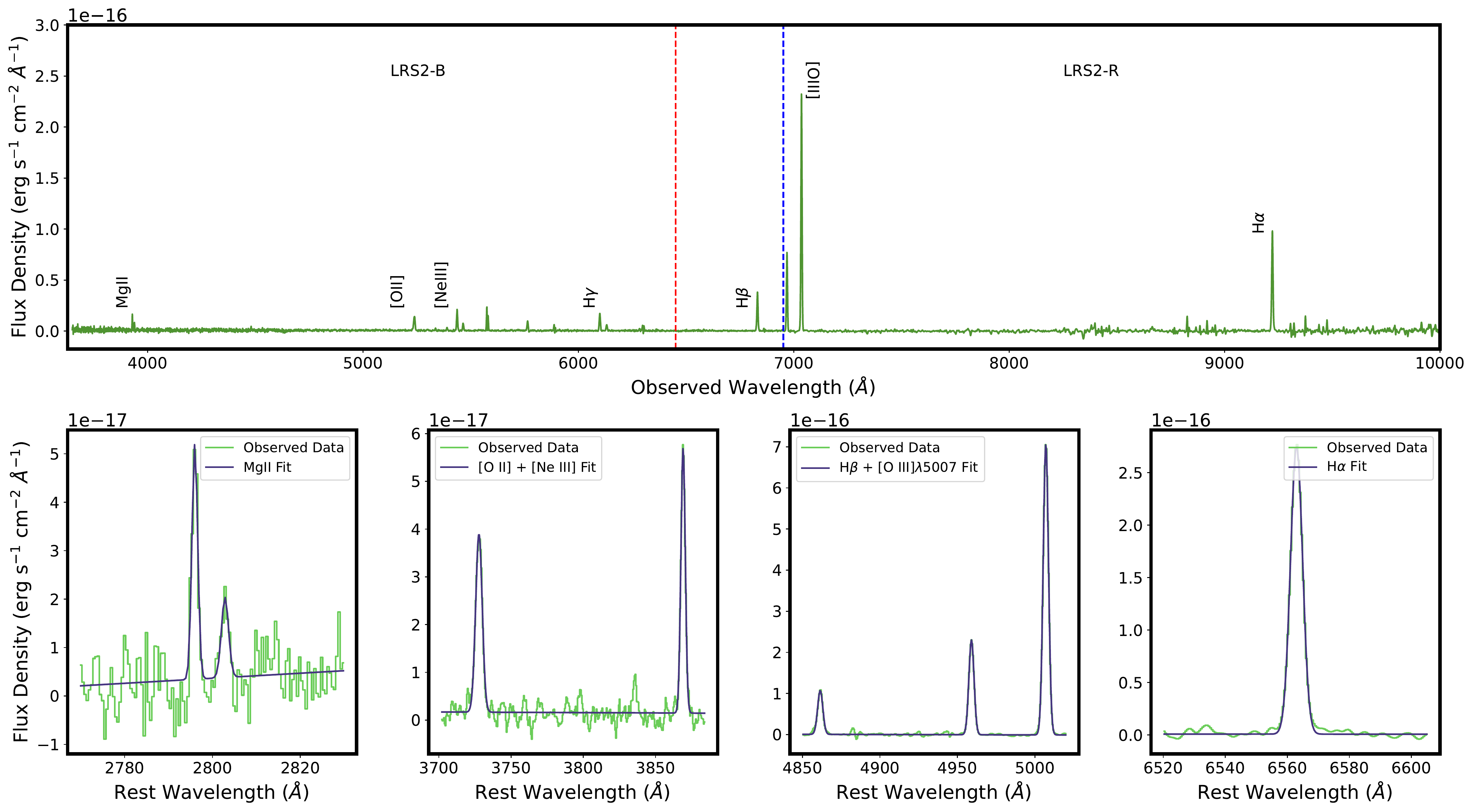}
    \caption{\textit{Top}: the total observed spectra from the Integrated aperture with a few lines of interest labelled. The blue dashed line marks the end of LRS2-B and the red dashed line marks the beginning of LRS2-R. \textit{Bottom}: A zoom in on some emission line fits with the restframe observed spectra in green and the fit in purple. The lines that we fit are: \ion{Mg}{II}, [\ion{O}{II}]$\lambda\lambda3727,3729$, [\ion{Ne}{III}]$\lambda3869$, H$\gamma$, H$\beta$, [\ion{O}{III}]$\lambda4959$, [\ion{O}{III}]$\lambda5007$, and H$\alpha$.}
    \label{fig:fitexamples_figure}
\end{figure*}

We tabulated the results and took the mean and standard deviation of the distribution. These techniques allowed us to measure the properties and errors of the emission lines of interest in a consistent way. Examples of our fits for the spectrum from the integrated region can be found in \autoref{fig:fitexamples_figure}. \autoref{tab:all_fluxes_table} gives the observed and extinction corrected fluxes (\autoref{sub:dust}), respectively, for 9 different measured emission lines in our 6 different apertures. 

All conversions from wavelengths to velocities were done using the restframe wavelengths from the NIST Atomic Spectra Database Lines Form \citep{NIST_ASD}. The lines measured in this work were \ion{Mg}{II}$\lambda2796,2803$, [\ion{O}{II}]$\lambda\lambda3727,3729$, [\ion{Ne}{III}]$\lambda3869$, H$\gamma$, [\ion{O}{III}]$\lambda4363$, H$\beta$, [\ion{O}{III}]$\lambda4959$, [\ion{O}{III}]$\lambda5007$, and H$\alpha$.

\begin{figure}
	\includegraphics[width=\columnwidth]{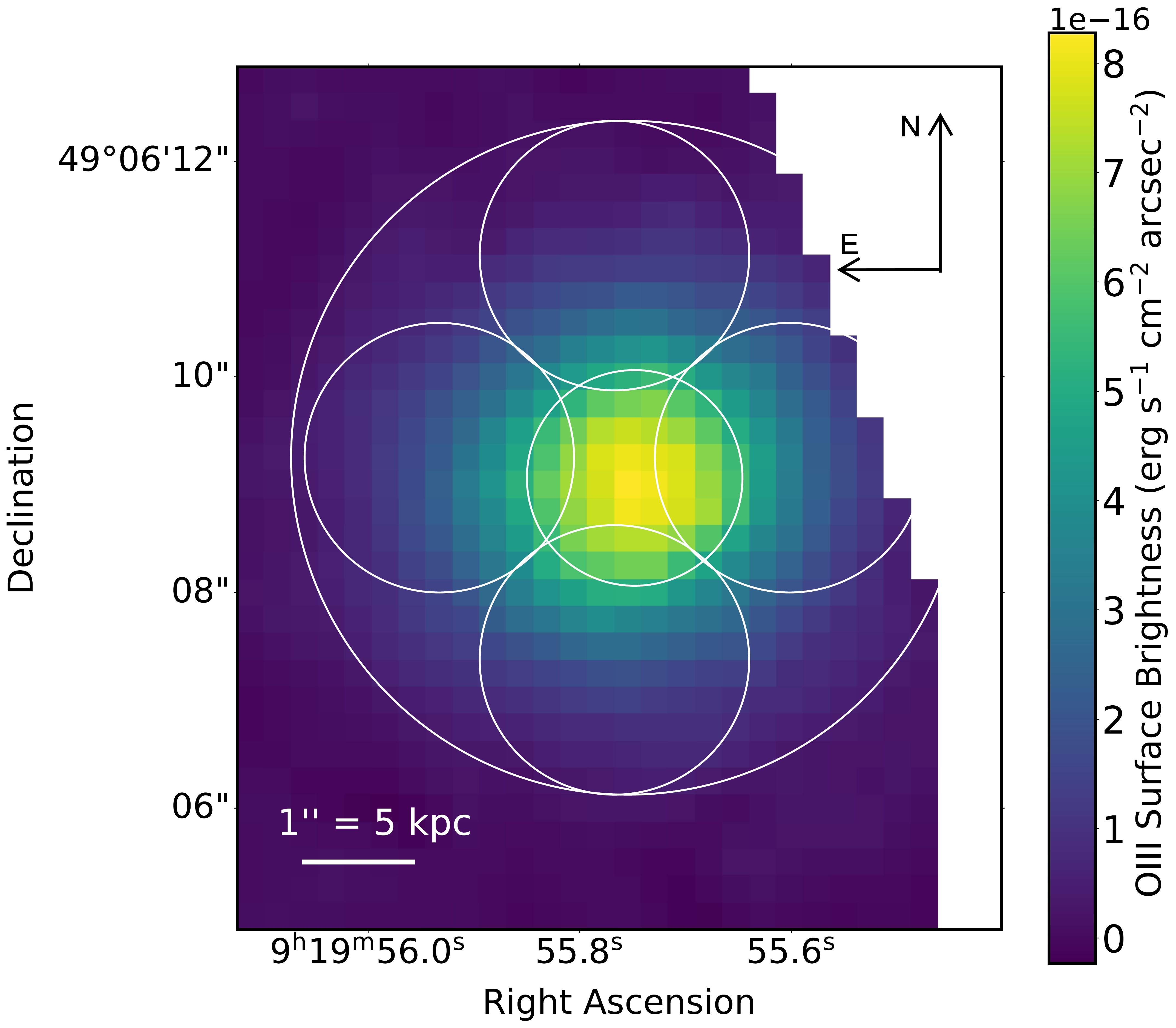}
    \caption{The continuum subtracted surface brightness spatial map for [\ion{O}{iii}]$\lambda5007$. The largest aperture is referred to as the Integrated aperture and the others are named after their position relative to the galaxy center, as indicated by the directional arrows (e.g. the aperture at the top of the image is the Northern aperture, the one on the left is the Eastern aperture, etc.). The blank spaxels are due to the object being observed on the edge of the detector. We include a 1~${\rm arcsec}$ scale bar, where 1~${\rm arcsec}$ corresponds to approximately 5 kpc in the frame of the galaxy.}
    \label{fig:aperture_figure}
\end{figure}

\begin{figure*}
	\includegraphics[width=\textwidth]{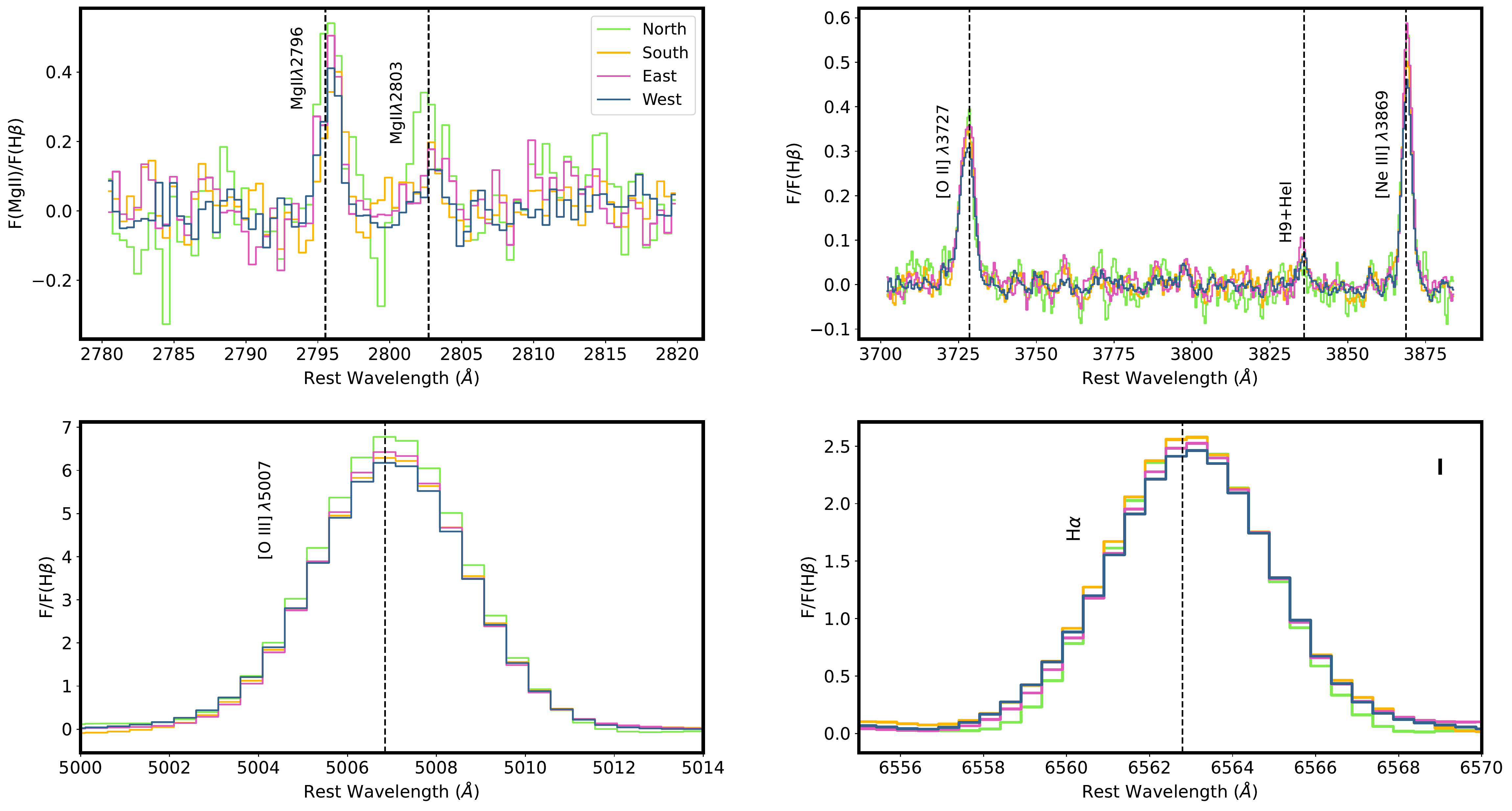}
    \caption{The restframe spectra of  \ion{Mg}{II}, [\ion{O}{II}]$\lambda\lambda3727,3729$, [\ion{Ne}{III}]$\lambda3869$, H$\beta$, [\ion{O}{III}]$\lambda5007$, and H$\alpha$ from each aperture, overlaid on one another. The flux densities were continuum subtracted but not dust extinction corrected. The flux densities were normalized by the maximum H$\beta$ flux density in their respective aperture such that all flux densities are relative to H$\beta$. The vertical dashed lines represent the rest wavelength line center. The bold line in the upper right corner of the bottom right panel represents the mean normalized error of the spectra.}
    \label{fig:overlay_figure}
\end{figure*}

\subsection{Comparison of fluxes and ratios to previous work}
\label{sub:sdss check}
To compare the LRS2 observations presented here to other literature measurements, we extracted the LRS2 flux from a 2~${\rm arcsec}$ diameter aperture, centered on the peak emission within J0919 (called the central aperture; the center most aperture in \autoref{fig:aperture_figure}). This aperture matches the diameter of the BOSS fibers \citep{boss_2013}. We then downloaded the calibrated spectra from the Sloan Digital Sky Survey DR15 \citep{SDSS_2019} and measured the emission line properties in the same way as was done in Sections \ref{sub:cont_fit}-\ref{sub:dust} with the LRS2 data. Our values from the SDSS spectra match literature values for integrated flux, equivalent width, and E(B-V) within $1\sigma$ for all emission lines (\citealt{Izotov_2021}, \citealt{Flury_2022}).

\autoref{tab:comparisons_table} compares the emission line ratios measured from the LRS2 (top row) and SDSS (bottom row) spectra. Most of the values measured from LRS2 match what we measured from the SDSS. For example, we measured a value of 3.04~$\pm$~0.02 for $\frac{{\rm H}\alpha}{{\rm H}\beta}$ from the LRS2 data. This is consistent, at the 2$\sigma$ significance level, with the value we measured from the SDSS data of 2.97~$\pm$~0.03. Our value for the [\ion{O}{III}]$\lambda5007$/[\ion{O}{II}]$\lambda\lambda3727,3729$ ratio is also consistent with the SDSS measurement. 



We used a redshift value of 0.40512 from the SDSS in our calculations. We found that the LRS2 emission lines have values of (24-37) km~s$^{-1}$ from this redshift. With our measurements matching other independent measurements, we can move forward assuming accurate results from our analysis. 

\subsection{Dust extinction correction}
\label{sub:dust}
We corrected the continuum subtracted emission-line flux values to account for the impact of dust present in the Milky Way (MW) and J0919. Dust extinction reduces the amount of flux that reaches our telescope by absorbing and/or scattering the photons of interest. It is also wavelength dependent meaning very blue lines, like [\ion{O}{II}]$\lambda\lambda3727,3729$, are more reduced than redder lines like H$\beta$. Correcting for dust extinction and comparing to uncorrected values constrains the spatial distribution of dust and reveals the intrinsic nebular conditions (e.g. metallicity, ionization structure, etc.). The correction was done using the following steps:
\begin{enumerate}
    \item Corrected the flux based on J0919's position in the Milky Way (MW) by multiplying the flux by: \begin{equation}
        10^{0.4 \times {\rm E(B-V)}_{\rm MW} \times k(\lambda_{\rm obs})}, 
    \end{equation}
    where {\rm E(B-V)}$_{\rm MW}$ is the MW color excess, which has a value of $0.029$ at the position of J0919 \citep{Green_2019}, and $k$($\lambda_{\rm obs}$) is the value of the CCM89 extinction law at the observed wavelength of each individual emission line \citep{CCM_1989}.
    \item Calculated the color excess intrinsic to J0901, referred to as E(B-V), by following the steps outlined in section 5.1 of \citealt{Flury_2022}.
    In short, we determine the variance-weighted E(B-V) by using the equivalent widths and the fluxes of ${\rm H}\alpha$, ${\rm H}\beta$, ${\rm H}\gamma$, and ${\rm H}\delta$. The E(B-V) value and stellar absorption values are iterated until the electron temperature (see \autoref{sub:metallicities}) converges. The E(B-V) values calculated using the stellar absorption are only used if the stellar absorption values are statistically significant at greater than the 2$\sigma$ level. This only applies to the Integrated and Western apertures. Uncertainties in the dust are folded into our extinction corrected flux errors. All other E(B-V) values are calculated with only the Balmer decrements. The values assumed for the intrinsic Balmer ratios depend on the temperature of the region, but given an average temperature of 16740 K they are; ${\rm H}\alpha$/${\rm H}\beta$: 2.77, ${\rm H}\gamma$/${\rm H}\beta$: 0.47 , ${\rm H}\delta$/${\rm H}\gamma$: 0.56. 
    \item To correct for dust in J0919 we multiplied the MW corrected fluxes by:
    \begin{equation}
    \label{eq:gal_correction}
        10^{0.4 \times {\rm E(B-V)} \times  k(\lambda_{\rm rest})}
    \end{equation}
\end{enumerate}
where $k$($\lambda$\textsubscript{rest}) is the value of the CCM89 extinction law at the rest wavelength of each individual emission line. We did not correct the flux values in the [\ion{O}{iii}]$\lambda5007$ spatial map (see \autoref{fig:aperture_figure}) in order to retain the observed spatial extent of J0919. However, we did apply these corrections to all the rows marked with "Cor" in \autoref{tab:all_fluxes_table}. This was done to highlight spatial differences between the apertures and to prepare the flux values to calculate the metallicites and escape fractions (see \autoref{sub:metallicities} and \autoref{sub:IDE}).
\subsection{Determining metallicities}
\label{sub:metallicities}
For each aperture we used \textsc{PyNeb} and the extinction-corrected [\ion{O}{III}]$\lambda5007$, [\ion{O}{III}]$\lambda4363$, [\ion{O}{II]}$\lambda\lambda3727,3729$ fluxes, all normalized by the H$\beta$ flux, to calculate the Oxygen abundances using the direct-T$_e$ method \citep{Garnett_1992,Berg_2013}. To determine the electron temperature, we used the temperature sensitive ratio of [\ion{O}{III}]$\lambda5007$ to the auroral [\ion{O}{III}]$\lambda4363$ line. Our calculated electron temperatures range from 15280-18200 K. We then used the [\ion{O}{II}]$\lambda\lambda3727,3729$ and [\ion{O}{III}]$\lambda5007$ fluxes, relative to H$\beta$, to determine the oxygen abundances in the intermediate (for [\ion{O}{II}]$\lambda\lambda3727,3729$) and high-ionization (for [\ion{O}{III}]$\lambda5007$) zone by assuming a single temperature across the \ion{H}{ii} region. By considering the total oxygen abundance as the sum of the intermediate and high ionization zones, we calculated the total oxygen abundance. This is a good approximation for galaxies that are as highly ionized as our target.

We detect the [\ion{O}{III}]$\lambda4363$ line at $>4\sigma$ significance in all of our apertures and list the inferred metallicity from each aperture in the last column of \autoref{tab:ratios_table} as 12+log(O/H). \citet{Izotov_2021} calculated an electron temperature of 16660~$\pm$~1440 and a metallicity of 7.77~$\pm$~0.01. The 12+log(O/H) value is broadly consistent with our measurement in the central aperture (\autoref{tab:ratios_table}).

\section{Results}
\label{sec:results}
In the following subsections we explore the properties of J0919, compare ionization ([\ion{O}{iii}]$\lambda5007$/[\ion{O}{II}]$\lambda\lambda3727,3729$) to dust (E(B-V)) and compare ionization to \ion{Mg}{II}$\lambda$2796. While our 6 apertures represent too small of a sample to be statistical, this analysis aims to quantify the spatial variation of \ion{Mg}{II} within a single LyC emitting galaxy, at z$\approx$0.4, to assess the spatial variation of the LyC escape.

\subsection{Integrated and resolved galaxy properties}
The global properties of J0919 are measured in the Integrated aperture (see \autoref{fig:fitexamples_figure}). From \autoref{tab:all_fluxes_table}, the observed integrated \ion{Mg}{II} flux (\ion{Mg}{II}$\lambda2796$ plus \ion{Mg}{II}$\lambda2803$) is 1.2$\times 10^{-16}$  erg s$^{-1}$ cm$^{-2}$, which corresponds to a total \ion{Mg}{II} luminosity of 7.36$\times 10^{40}$ erg s$^{-1}$ at z$\approx$0.4. These values are 3 times larger than those measured in the Central aperture and by extension, indicate that the SDSS data considered in \autoref{sub:sdss check} may not capture all of the flux from J0919. The other values of interest from the Integrated aperture, found in \autoref{tab:ratios_table}, are: [\ion{O}{III}]$\lambda5007$/[\ion{O}{II}]$\lambda\lambda3727,3729$ = 12.4~$\pm$~0.2; H$\alpha$/H$\beta$ = 2.98~$\pm$~0.04; 12+log(O/H) = 7.807~$\pm$~0.006; and E(B-V) = 0.072~$\pm$~0.008. As with the \ion{Mg}{II} flux, these Integrated values are different from the Central values. We compare the values from the various apertures in \autoref{sec:discussion}.

On a smaller spatial scale than the Integrated aperture, \autoref{fig:overlay_figure} overlays the H$\beta$-normalized spectra of the spatially distinct apertures. The H$\beta$ normalization puts the apertures on a common scaling so that they can be compared. From this comparison a visual difference can be seen between the different spectra; differences which indicate that the emission features vary per region. If the physical conditions were the same in all the regions, the spectra would look identical when normalized. Specifically in the case of H$\alpha$/H$\beta$ (bottom right panel \autoref{fig:overlay_figure}), which is a component of the dust attenuation calculation (see \autoref{sub:dust}), the spectra are not identical and span values of (2.79~$\pm$~0.04)-(3.28~$\pm$~0.07). This range of H$\alpha$/H$\beta$ spans from completely dust-free to moderately reddened. Similarly significant differences can be seen between the apertures in the \ion{Mg}{II} flux (upper left panel \autoref{fig:overlay_figure}) where the Northern aperture has significantly more flux than the Western aperture, for example. The differences in \autoref{fig:overlay_figure} are also seen in the \ion{Mg}{II}/[\ion{O}{III}]$\lambda5007$ (0.013-0.024), [\ion{O}{III}]$\lambda5007$/[\ion{O}{II}]$\lambda\lambda3727,3729$ (10.9-14.1), and [\ion{Ne}{III}]$\lambda3869$/[\ion{O}{II}]$\lambda\lambda3727,3729$ (0.97-1.04) emission features. Thus, this figure serves as the first hint for spatial variations. We will investigate spatial variations in terms of dust and ionization properties in the next subsections.

\subsection{Relationship between ionization and dust}
\begin{figure}
	\includegraphics[width=\columnwidth]{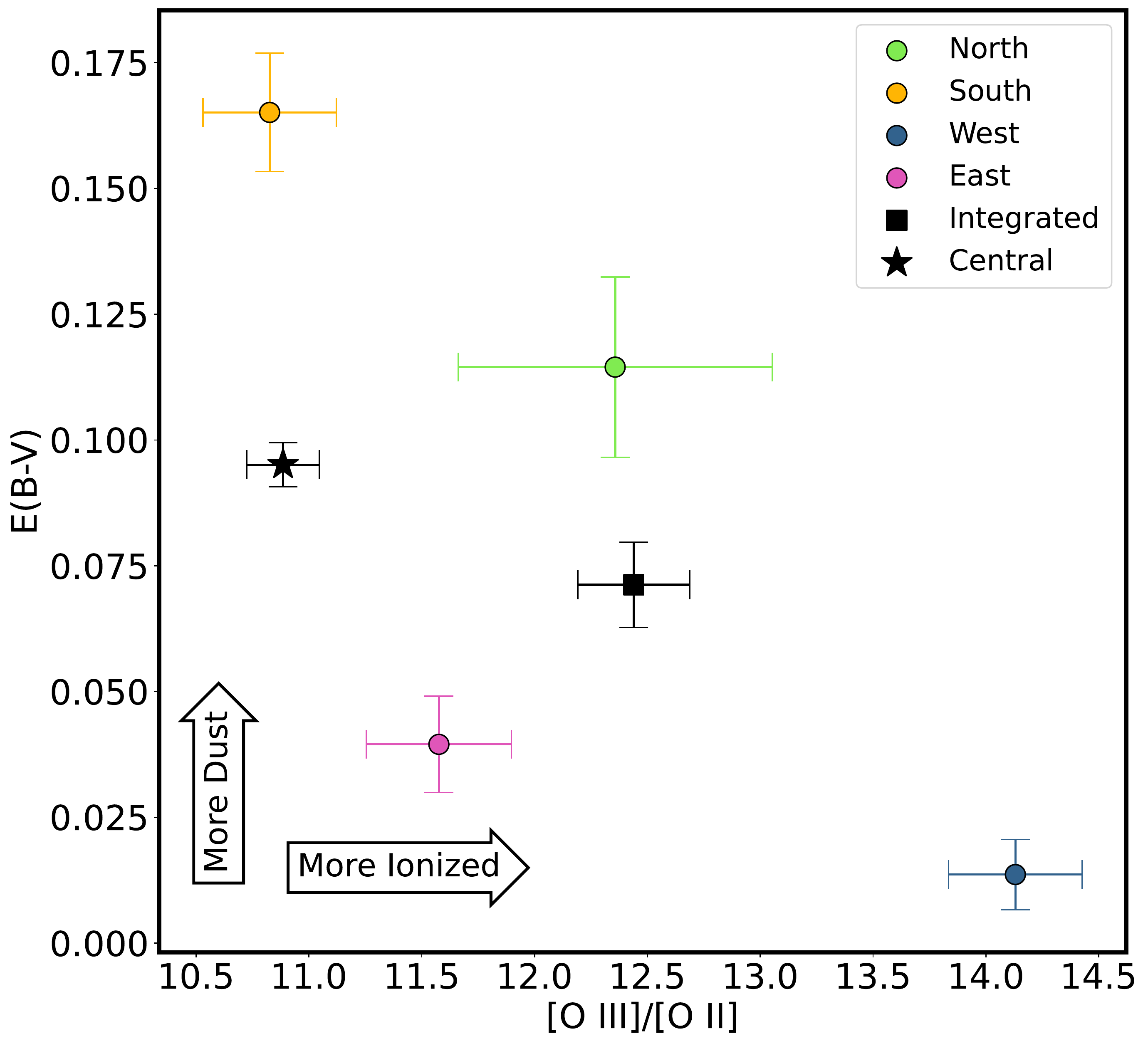}
    \caption{The E(B-V) of the individual apertures versus the [\ion{O}{III}]$\lambda5007$/[\ion{O}{II}]$\lambda\lambda3727,3729$ ratio. The E(B-V) value traces dust attenuation and the [\ion{O}{III}]$\lambda5007$/[\ion{O}{II}]$\lambda\lambda3727,3729$ ratio traces ionization. Dust content increases as E(B-V) increases above 0 (see arrows). The legend caption refers to the aperture location.}
    \label{fig:o32vdust_figure}
\end{figure}

Within the frame of understanding local LyC escape, we explored the relationship between ionization and dust in each aperture. Ionization is important for LyC escape because galaxies that are more ionized have less relative neutral hydrogen. We use the [\ion{O}{III}]$\lambda5007$/[\ion{O}{II}]$\lambda\lambda3727,3729$ ratio because this ratio directly traces the fraction of highly ionized to moderately ionized gas in a metallicity independent way. As such it is a diagnostic of the ionization state of the gas, with higher [\ion{O}{III}]$\lambda5007$/[\ion{O}{II}]$\lambda\lambda3727,3729$ values corresponding to more highly ionized nebulae. Dust preferentially absorbs and scatters bluer wavelengths, meaning that it strongly absorbs ionizing photons. This relationship is shown in \autoref{fig:o32vdust_figure}. The dust values are represented by E(B-V), where a value of 0 represents a dust free nebula (see \autoref{sub:dust}). The [\ion{O}{III}]$\lambda5007$ and [\ion{O}{II}]$\lambda\lambda3727,3729$ fluxes were corrected for dust attenuation because we were interested in the intrinsic ionization of the aperture.

We measure different ionization and dust attenuation values for each aperture and find spatial variations between the apertures (\autoref{fig:o32vdust_figure}). To verify that we are seeing the effects of spatial variation and not a dependence on dust attenuation for both axes, we calculated [\ion{O}{III}]$\lambda5007$/[\ion{O}{II}]$\lambda\lambda3727,3729$ values using the E(B-V) $\pm 3\sigma$ from the Integrated aperture. We found that the ratio ranged between 12.1~$\pm$~0.3 and 12.8~$\pm$~0.3. This range is too small to explain the variation seen in \autoref{fig:o32vdust_figure}. We also find a negative trend between ionization and dust, where dust-free regions are more ionized, as expected.




\subsection{Relationship between dust and \ion{Mg}{II}$\lambda$2796 flux}
\label{sub:dust mgii}
There are two sinks for ionizing photons: dust and neutral gas. One of our goals is to explore the relative impact of both dust and neutral gas on a spatial basis within J0919. We can explore neutral gas using the \ion{Mg}{II} doublet ratio or photoionization models \citep{Henry_2018,Chisholm_2020}. The two \ion{Mg}{II} transitions (\ion{Mg}{II}$\lambda2796$ and \ion{Mg}{II}$\lambda2803$) have different oscillator strengths. This means that their observed relative flux ratios are sensitive to the \ion{Mg}{II} column density. However, the SNR of \ion{Mg}{II}$\lambda2803$ averages 3 in individual apertures, which is too low to allow for an analysis using the doublet ratio technique on a spatially resolved basis (see \autoref{tab:all_fluxes_table}). Because we do not use \ion{Mg}{II}$\lambda2803$, any mention of \ion{Mg}{II} in the rest of the paper will be referring to \ion{Mg}{II}$\lambda2796$ exclusively. We note, however, that the total integrated flux ratio that we measured (2.1±0.6, see \autoref{tab:ratios_table}) is broadly consistent with the results presented here and suggests that \ion{Mg}{II} photons are on average escaping in an optically thin medium \citep{Chisholm_2020}. In place of the doublet ratio we take the observed \ion{Mg}{II}$\lambda2796$ values and normalize them by the dust attenuation corrected [\ion{O}{III}]$\lambda5007$ values. The fraction of \ion{Mg}{II}$\lambda2796$ emission we observe relative to the [\ion{O}{III}] emission is related to the neutral gas column density \citep{Henry_2018}. The relationship between neutral gas column density and ionization is presented in the next subsection.

\begin{figure}
	\includegraphics[width=\columnwidth]{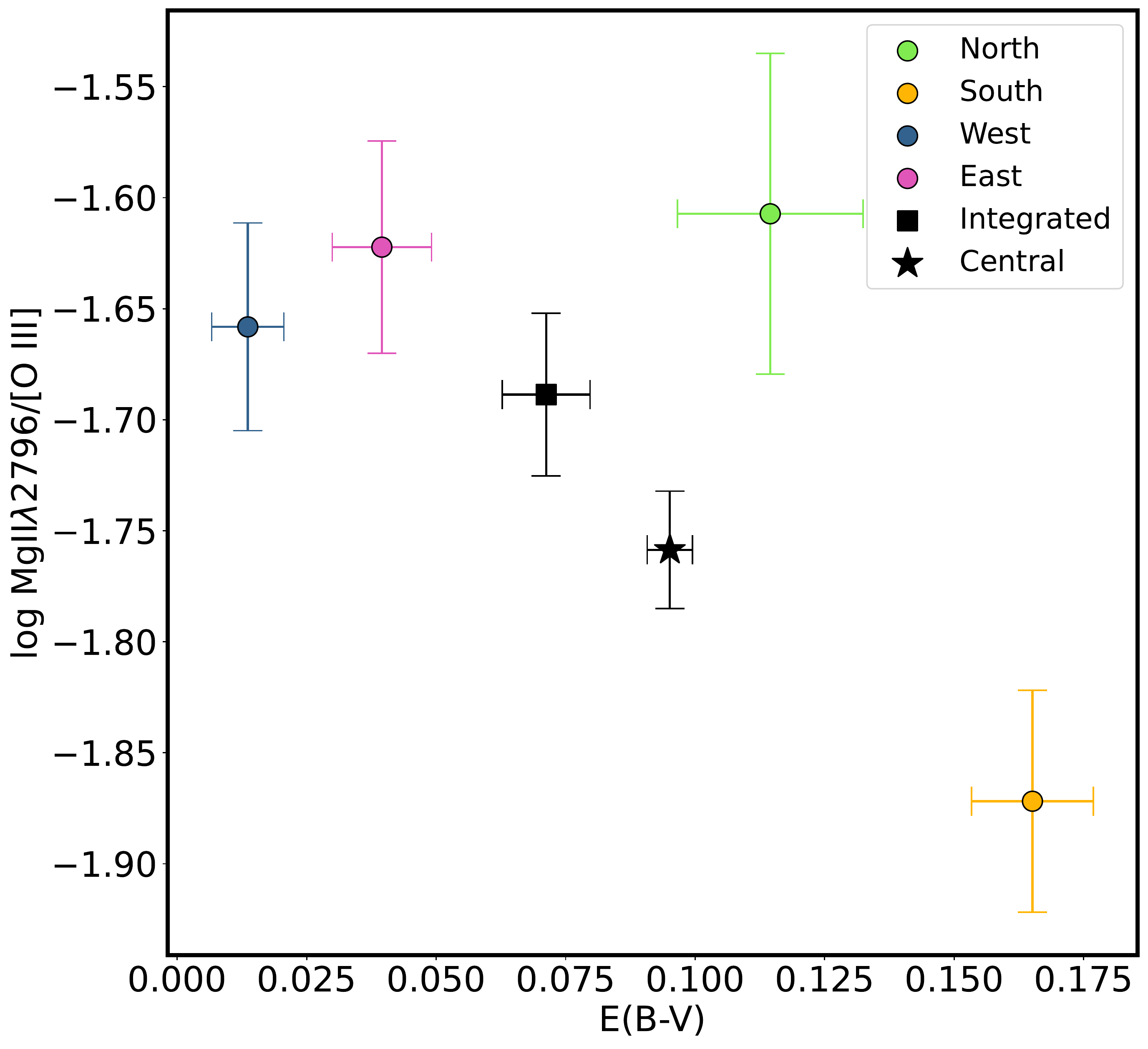}
    \caption{The E(B-V) of the different apertures versus the observed \ion{Mg}{II}$\lambda2796$ flux, normalized by the dust attenuation corrected [\ion{O}{III}]$\lambda5007$ flux. The E(B-V) value traces dust attenuation and has a value of 0 in the absence of dust. Generally, regions within J0919 with the highest observed \ion{Mg}{II}/[\ion{O}{III}]$\lambda5007$ ratios also have the lowest dust attenuation. The legend caption refers to the aperture location within J0919.}
    \label{fig:dustvmgii_figure}
\end{figure}

We measure significant \ion{Mg}{II}$\lambda2796$ flux variation for each aperture (\autoref{tab:all_fluxes_table}). The \ion{Mg}{II}$\lambda2796$ line, normalized by the [\ion{O}{III}]$\lambda5007$ line, is negatively correlated with E(B-V) (\autoref{fig:dustvmgii_figure}). More \ion{Mg}{II}$\lambda2796$ relative to [\ion{O}{III}] emission escapes the galaxy in less dusty regions of J0919. In \autoref{sub:IDE} we use photoionization models to explore how this observation relates to the escape of ionizing photons.

\subsection{Relationship between ionization and \ion{Mg}{II}$\lambda$2796 flux}
\label{sub:ionvsmgii}

The ionization state of the gas traces the amount of low ionization relative to high ionization gas. In regions with high ionization there is less neutral gas that can absorb LyC photons and the \ion{Mg}{II}$\lambda2796$/[\ion{O}{III}]$\lambda5007$ ratio allows for a comparison between the observed \ion{Mg}{II} values and the predicted values from the photoionization models of \citet{Henry_2018} (see \autoref{sub:IDE}).

\begin{figure}
	\includegraphics[width=\columnwidth]{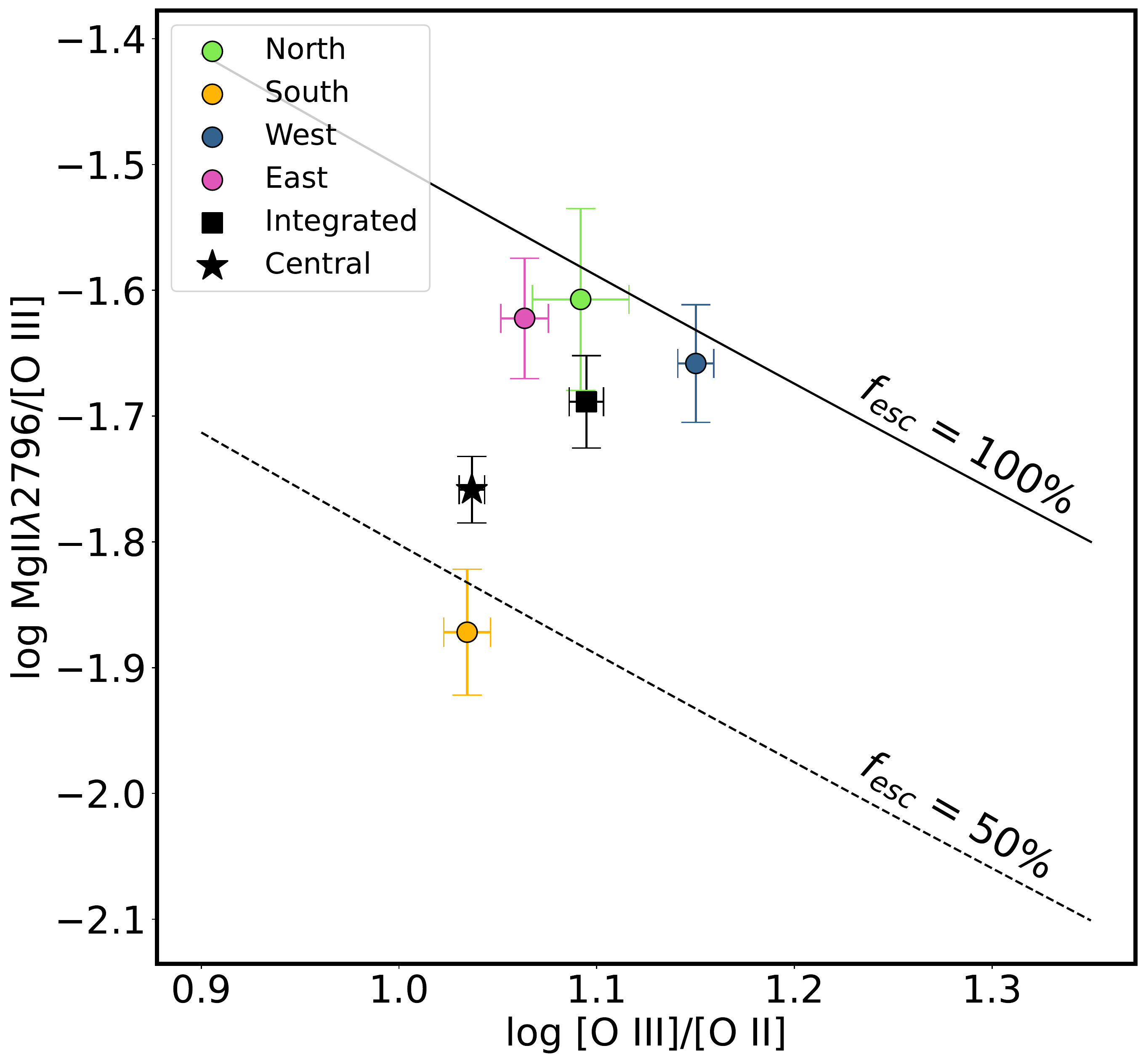}
    \caption{The ionization of the galaxy measured with the dust attenuation corrected [\ion{O}{III}]$\lambda5007$/[\ion{O}{II}]$\lambda\lambda3727,3729$ ratio versus the observed \ion{Mg}{II}$\lambda2796$ flux, normalized by the dust attenuation corrected [\ion{O}{III}]$\lambda5007$ flux. The lines represent different \ion{Mg}{II}$\lambda$2796 escape fractions \citep{Henry_2018}. Regions within J0919 with the highest ionization have the largest \ion{Mg}{II} escape fractions. The legend caption refers to the aperture location.}
    \label{fig:o32vmgii_figure}
\end{figure}

In each of our apertures, we measure different ionization and \ion{Mg}{II}$\lambda2796$ flux values. \autoref{fig:o32vmgii_figure} shows a weak positive trend between ionization, measured with the dust attenuation corrected [\ion{O}{III}]$\lambda5007$/[\ion{O}{II}]$\lambda\lambda3727,3729$ ratio, and the observed \ion{Mg}{II}$\lambda$2796 flux, which was normalized by the dust attenuation corrected [\ion{O}{III}]$\lambda5007$ flux. 

This indicates that regions within J0919 that have the highest \ion{Mg}{II} relative to [\ion{O}{III}] emission are also the most ionized. We find similar trends with the [\ion{Ne}{III}]$\lambda3869$/[\ion{O}{II}]$\lambda3728$ flux ratios (see \autoref{tab:ratios_table}) but use the [\ion{O}{III}]$\lambda5007$/[\ion{O}{II}]$\lambda3728$ ratio because [\ion{O}{III}]$\lambda5007$ has significantly higher SNR than [\ion{Ne}{III}]$\lambda3869$.


\begin{figure}
	\includegraphics[width=\columnwidth]{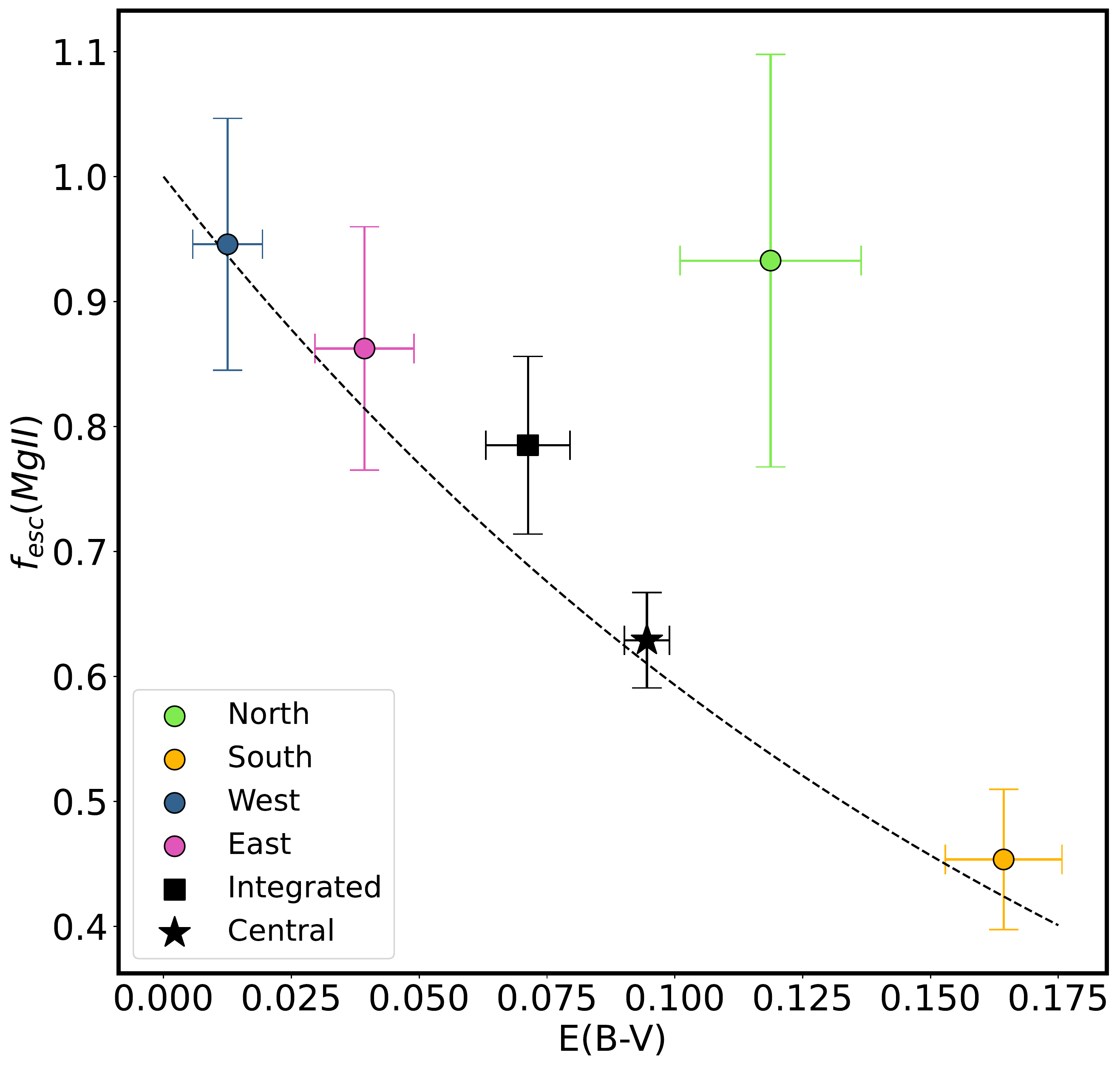}
    \caption{The escape fraction of \ion{Mg}{II} of the different apertures versus the E(B-V). The E(B-V) value traces dust attenuation and has a value of 0 in the absence of dust. Generally, regions within J0919 with the highest observed f\textsubscript{esc}(\ion{Mg}{II}) also have the lowest dust attenuation. The dotted line traces the theoretical relationship between f\textsubscript{esc}(\ion{Mg}{II}) and E(B-V). The legend caption refers to the aperture location within J0919.}
    \label{fig:fescvebmv_figure}
\end{figure}

\begin{table*}
    \centering
    \caption{E(B-V) values, attenuation corrected integrated fluxes, and observed integrated fluxes for 6 different apertures and 9 different emission lines. The fluxes are in units of ($10^{-17}$  erg s$^{-1}$ cm$^{-2}$). E(B-V) is in units of magnitudes. Cor indicates that the values have been attenuation corrected. Abs indicates that the Balmer lines have been corrected for stellar absorption. If the row only has the Abs label, the Abs column will contain the value for stellar absorption used in the correction in \AA\ (see \citealt{Izotov_1994} for more details). [\ion{O}{II}] refers to [\ion{O}{II}]$\lambda\lambda3727,3729$. The aperture naming convention refers to the aperture's position relative to the galaxy's center (see legend of \autoref{fig:overlay_figure}) where the Integrated aperture is the largest aperture in \autoref{fig:aperture_figure}.} 
    \label{tab:all_fluxes_table}
    \begin{tabular}{lcccccccccr}
        \hline
        Region     & E(B-V)  & Abs   & \ion{Mg}{II}$\lambda2796$  & \ion{Mg}{II}$\lambda2803$  & [\ion{O}{II}] & [\ion{Ne}{III}]$\lambda3869$  & [O III]$\lambda4363$ & H$\beta$   & [OIII]$\lambda5007$ & H$\alpha$ \\ \hline
        Int. & -     & -           & 8.2~$\pm$~0.7  & 4~$\pm$~1  & 21.4~$\pm$~0.4  & 21.2~$\pm$~0.3   & 6.2~$\pm$~0.3   & 45.0~$\pm$~0.6  & 298.8~$\pm$~0.5 & 134.2~$\pm$~0.7  \\
        Int.\textsuperscript{Abs} & - & 27~$\pm$~3 & -   & -    & -       & -        & -   & 52~$\pm$~2   & - & 149~$\pm$~3 \\
        Int.\rlap{\textsuperscript{Cor}}\textsubscript{Abs} & 0.072~$\pm$~0.008 & - & 13~$\pm$~1   & 8~$\pm$~5    & 32~$\pm$~1   & 31~$\pm$~1  & 8.8~$\pm$~0.5   & 62~$\pm$~2   & 400~$\pm$~11 & 167~$\pm$~3 \\ 
        \\
        North        & -     & -   & 1.4~$\pm$~0.2 & 1~$\pm$~0.2 & 2.4~$\pm$~0.1   & 2.59~$\pm$~0.08 & 0.81~$\pm$~0.08  & 4.9~$\pm$~0.1  & 35.7~$\pm$~0.2 & 14.9~$\pm$~0.2     \\
        North\textsuperscript{Cor}       & 0.12~$\pm$~0.02 & - & 3.0~$\pm$~0.5   & 2.2~$\pm$~0.8 & 4.5~$\pm$~0.4    & 4.7~$\pm$~0.4    & 1.4~$\pm$~0.2   & 7.8~$\pm$~0.5    & 56~$\pm$~3 & 20.2~$\pm$~0.9    \\ 
        \\
        South     & -        & -       & 1.5~$\pm$~0.2 & 1~$\pm$~0.9 & 4.6~$\pm$~0.1  & 4.60~$\pm$~0.09  & 1.5~$\pm$~0.1 & 9.1~$\pm$~0.2  & 63.0~$\pm$~0.1 & 29.9~$\pm$~0.2   \\
        South\textsuperscript{Cor}    & 0.16~$\pm$~0.01 & - & 4.3~$\pm$~0.5   & 3.1~$\pm$~2.6   & 10~$\pm$~0.6    & 10~$\pm$~0.5    & 3.1~$\pm$~0.3    & 16.8~$\pm$~0.7    & 113~$\pm$~4 & 46~$\pm$~1  \\ 
        \\
        West  & -     & -        & 1.9~$\pm$~0.2 & 0.6~$\pm$~0.3 & 5.3~$\pm$~0.1   & 5.15~$\pm$~0.07  & 1.54~$\pm$~0.06  & 12.4~$\pm$~0.2  & 77.6~$\pm$~0.2 & 35.6~$\pm$~0.2  \\
        West\textsuperscript{Abs} & - & 41~$\pm$~8  & -   & -    & -       & -        & -   & 13.0~$\pm$~0.2   & - & 36.5~$\pm$~0.3 \\
        West\rlap{\textsuperscript{Cor}}\textsubscript{Abs}     & 0.013~$\pm$~0.007 & - & 2.3~$\pm$~0.3       & 1.7~$\pm$~0.6       & 6.1~$\pm$~0.2     & 5.9~$\pm$~0.2   & 1.74~$\pm$~0.08   & 14.4~$\pm$~0.4   & 86~$\pm$~2 & 39.1~$\pm$~0.7  \\ 
        \\
        East       & -      & -          & 1.7~$\pm$~0.2  & 1~$\pm$~0.7 & 4.7~$\pm$~0.1   & 4.70~$\pm$~0.08  & 1.20~$\pm$~0.09  & 9.5~$\pm$~0.1  & 58.2~$\pm$~0.2 & 26.4~$\pm$~0.2 \\
        East\textsuperscript{Cor}      & 0.04~$\pm$~0.01 & -  & 2.3~$\pm$~0.3    & 2~$\pm$~1  & 6.04~$\pm$~0.31    & 6.02~$\pm$~0.31   & 1.5~$\pm$~0.1    & 11.5~$\pm$~0.4   & 70~$\pm$~2 & 30.0~$\pm$~0.8   \\ 
        \\
        Central    & -   & -    & 2.7~$\pm$~0.2 & 0.9~$\pm$~0.2 & 8.5~$\pm$~0.1  & 8.71~$\pm$~0.09  & 2.84~$\pm$~0.08  & 17.5~$\pm$~0.1 & 107.7~$\pm$~0.1 & 17.5~$\pm$~0.1  \\
        Central\textsuperscript{Cor}   & 0.093~$\pm$~0.004 & -  & 5.1~$\pm$~0.3   & 1.8~$\pm$~0.3 & 14.1~$\pm$~0.3   & 14.2~$\pm$~0.3   & 4.4~$\pm$~0.2   & 25.4~$\pm$~0.4    & 153~$\pm$~2 & 68.7~$\pm$~0.7  \\ 
        \hline 
    \end{tabular}
\end{table*}

\begin{table}
    \centering
    \caption{Values from our measurements in the LRS2 2~${\rm arcsec}$ aperture and the SDSS spectra. The 2~${\rm arcsec}$ aperture is the central aperture in \autoref{fig:aperture_figure}. No fluxes have been dust corrected and the E(B-V) values were calculated using \autoref{sub:dust}}
    \label{tab:comparisons_table}
    \begin{tabular}{lcr}
        \hline
         Measurements  & LRS2 2~${\rm arcsec}$ & SDSS \\ \hline
        H$\gamma$/H$\beta$ & 0.464~$\pm$~0.005 & 0.44~$\pm$~0.01 \\ 
        E(B-V)    & 0.030~$\pm$~0.008 & 0.093~$\pm$~0.004 \\
        MgII2796/MgII2803 & 2.9~$\pm$~0.5 & 1.9~$\pm$~0.6 \\
        H$\alpha$/H$\beta$ & 3.04~$\pm$~0.02 & 2.97~$\pm$~0.03 \\\
        [O III]$\lambda5007$/[\ion{O}{II}]$\lambda\lambda3727,3729$ & 12.5~$\pm$~0.2 & 12.3~$\pm$~0.3 \\\
        [O III]$\lambda5007$/[O III]$\lambda4959$ & 3.11~$\pm$~0.01 & 3.06~$\pm$~0.02 \\\hline
    \end{tabular}
\end{table}


\begin{table*}
    \centering
    \caption{Emission line flux ratios for the 6 different apertures. All fluxes are dust extinction corrected except for the \ion{Mg}{II} and H$\alpha$/H$\beta$ ratios. [\ion{O}{II}] refers to [\ion{O}{II}]$\lambda\lambda3727,3729$.}
    \label{tab:ratios_table}
    \begin{tabular}{lcccccr}
        \hline
        Region     & \ion{Mg}{II}$\lambda2796$/\ion{Mg}{II}$\lambda2803$ & [\ion{O}{III}]$\lambda5007$/[\ion{O}{II}] & \ion{Mg}{II}$\lambda2796$/[\ion{O}{III}]$\lambda5007$ & [\ion{Ne}{III}]$\lambda3869$/[\ion{O}{II}] & H$\alpha$/H$\beta$ & 12+log(O/H)\\ \hline
        Integrated & 2.1~$\pm$~0.6    & 12.4~$\pm$~0.2       & 0.020~$\pm$~0.002    & 0.97~$\pm$~0.02       & 2.98~$\pm$~0.04  & 7.807~$\pm$~0.006\\ 
        North        & 1.4~$\pm$~0.4    & 12.3~$\pm$~0.7       & 0.024~$\pm$~0.004    & 1.04~$\pm$~0.06       & 3.05~$\pm$~0.08  & 7.78~$\pm$~0.01\\ 
        South     & 1.4~$\pm$~1.2    & 10.8~$\pm$~0.3       & 0.013~$\pm$~0.002    & 0.97~$\pm$~0.03       & 3.28~$\pm$~0.07  & 7.718~$\pm$~0.009\\ 
        West      & 3.1~$\pm$~1.7    & 14.1~$\pm$~0.3       & 0.022~$\pm$~0.002    & 0.97~$\pm$~0.02       & 2.87~$\pm$~0.04  & 7.817~$\pm$~0.006\\ 
        East       & 1.7~$\pm$~1.2    & 11.6~$\pm$~0.3       & 0.024~$\pm$~0.003    & 1.00~$\pm$~0.03       & 2.79~$\pm$~0.04  & 7.792~$\pm$~0.005\\ 
        Central    & 2.9~$\pm$~0.5    & 10.9~$\pm$~0.2       & 0.018~$\pm$~0.001    & 1.01~$\pm$~0.02       & 3.04~$\pm$~0.02  & 7.637~$\pm$~0.003\\ \hline
    \end{tabular}
\end{table*}

\begin{table}
    \centering
    \caption{Values from our measurements of the \ion{Mg}{II} and LyC escape fractions along with the ionizing photon flux estimate, Q\textsubscript{HI}, for the 6 individual apertures, a COS analogue aperture, and the average of the spatially distinct regions. All the escape fractions are percentages. F(H$\beta$) is the extinction corrected H$\beta$ flux. Q\textsubscript{HI} has units of ($10^{53}$ s$^{-1}$).}
    \label{tab:escapefractions}
    \begin{tabular}{lccr}
        \hline
        Region   & \ion{Mg}{II} Escape[\%] & LyC Escape[\%] & Q\textsubscript{HI} \\ \hline
        Integrated & 79~$\pm$~7    & 20~$\pm$~4 & 10.2~$\pm$~0.3\\ 
        North      & 94~$\pm$~16   & 10~$\pm$~5 & 1.14~$\pm$~0.04\\ 
        South     & 45~$\pm$~5    & 2.0~$\pm$~0.6 & 2.27~$\pm$~0.09 \\ 
        West      & 93~$\pm$~10    & 72~$\pm$~14 & 8~$\pm$~4\\ 
        East      & 87~$\pm$~10   & 42~$\pm$~11 & 2.6~$\pm$~0.4 \\ 
        Central   & 60~$\pm$~4    & 10~$\pm$~1  & 3.76~$\pm$~0.04 \\ 
        2.5~${\rm arcsec}$ (COS) & 63~$\pm$~4  & 13~$\pm$~2  & 5.17~$\pm$~0.05 \\
        Average    & 80~$\pm$~4    & 32~$\pm$~5  & 3~$\pm$~1 \\\hline
    \end{tabular}
\end{table}

\section{Discussion}
\label{sec:discussion}
In this section we comment on the variations in our different apertures using the quantities in Tables \ref{tab:all_fluxes_table} and \ref{tab:ratios_table}. We also discuss the implications of the trends between dust, ionization, and \ion{Mg}{II} from \autoref{sec:results} on the escape of \ion{Mg}{II} and LyC photons.

\subsection{Ionization, dust, and escape fractions}
\label{sub:IDE}
To better understand how \ion{Mg}{II} serves as a means to trace neutral H and the escape of ionizing photons, we study the spatial variation of \ion{Mg}{II}. Confirming methods to infer the LyC escape fraction locally, will allow future observations to determine how the distant universe was reionized. In Section \ref{sec:results}, we show that there are potential relations between the \ion{Mg}{II} emission, nebular ionization, and dust attenuation. These correlations suggest that a larger fraction of \ion{Mg}{II} escapes regions of higher ionization and lower dust attenuation (\autoref{fig:o32vdust_figure}, \autoref{fig:dustvmgii_figure}, \autoref{fig:o32vmgii_figure}, \autoref{fig:fescvebmv_figure}). Both of these conditions are consistent with \ion{Mg}{II} being a tracer of LyC escape because both dust and low-ionization (neutral) gas absorbs ionizing photons \citep{Chisholm_2020}. While these empirical trends suggest that more \ion{Mg}{II} photons, and by extension LyC photons, will escape highly ionized, dust-free regions, it does not tell us what fraction of \ion{Mg}{II} escapes these regions.

To calculate the \ion{Mg}{II} escape fractions, we used equation 1 from \cite{Henry_2018} and our dust attenuation corrected [\ion{O}{III}]$\lambda5007$/[\ion{O}{II}]$\lambda\lambda3727,3729$ ratio to calculate the intrinsic \ion{Mg}{II}$\lambda2796$/[\ion{O}{III}]$\lambda5007$ ratio. We calculated the \ion{Mg}{II}$\lambda2796$/[\ion{O}{III}]$\lambda5007$ flux ratio with the observed \ion{Mg}{II}$\lambda2796$ flux and the dust attenuation corrected [\ion{O}{III}]$\lambda5007$ flux. The ratio of our observed \ion{Mg}{II}/[\ion{O}{III}] flux ratio to the intrinsic ratio is our reported \ion{Mg}{II} escape fraction (see \autoref{tab:escapefractions}). Combining \autoref{fig:o32vdust_figure}, \autoref{fig:dustvmgii_figure}, and \autoref{fig:o32vmgii_figure}, we find that regions within J0919 with low-to-moderate dust attenuation that are highly ionized emit roughly 90\% of their intrinsic \ion{Mg}{II} emission (see \autoref{fig:fescvebmv_figure}). Regions that are lower ionization and higher dust attenuation emit $\sim$45\% of their intrinsic \ion{Mg}{II} emission. This factor of $\sim2$ difference in \ion{Mg}{II} escape fraction illustrates that there are strong and significant spatial variations in the \ion{Mg}{II} escape (see \autoref{sub:aperture variations}).

Finally, we follow eq. 25 from \citet{Chisholm_2020} to extend the \ion{Mg}{II} escape fraction to the LyC escape fraction by dust correcting the \ion{Mg}{II} escape fraction (\autoref{tab:escapefractions}). \citet{Chisholm_2020} found that the \ion{Mg}{II} escape fractions over-predicted the LyC escape fraction and required a further dust correction to match observed LyC escape fractions. To do this, we convert the E(B-V) inferred from the \citet{CCM_1989} attenuation curve to one using the \citet{Reddy_2016} reddening curve (see Appendix A of \citealt{Chisholm_2022}). We use the value of the \citet{Reddy_2016} attenuation law at 912\AA\ ($k$(912) = 12.87) to accomplish this. Since regions with high dust attenuation already have lower \ion{Mg}{II} escape fractions, these regions can have an order of magnitude lower LyC escape fractions than regions with higher \ion{Mg}{II} escape. \citet{Izotov_2021} measured a LyC escape fraction of 16\% along a single line of sight through the central portion of J0919. This value is consistent with the LyC escape fraction we infer from the COS analogue aperture (\autoref{tab:escapefractions}, 13\%). This confirms that \ion{Mg}{II} and dust are sufficient to reproduce the observed escape fraction of ionizing photons in galaxies \citep{Chisholm_2020}. The high dust attenuation and low ionization (more neutral gas) of these regions compound to reduce the LyC escape fractions. Different sightlines through the same galaxy are either transparent or opaque to ionizing photons.

In summary, we find that regions within J0919 that are highly ionized and have low dust attenuation emit a larger fraction of their intrinsic \ion{Mg}{II}, and by extension, LyC photons. The correlations found in Figures \ref{fig:o32vdust_figure}, \ref{fig:dustvmgii_figure}, and \ref{fig:o32vmgii_figure} illustrate that the conditions that are likely to lead to LyC escape are intimately intertwined: low-dust and highly ionized regions within galaxies may have the greatest amounts of \ion{Mg}{II} emission.

\subsection{Variations of indirect tracers between apertures}
\label{sub:aperture variations}
Recent observations have successfully found local galaxies that emit ionizing photons. These observations have found escape fractions between 0-73\%, with a large scatter in many of the classically expected diagnostics -- like [\ion{O}{III}]$\lambda5007$/[\ion{O}{II}]$\lambda\lambda3727,3729$ and H$\beta$ equivalent widths \citep{Naidu_2018,Fletcher_2019,Izotov_2021,Flury_2022b}. To reconcile these discrepancies with a complete understanding of reionization would require a measurement of the volume-averaged escape fraction from LyC leakers. However, all of the direct LyC detections are along a single sightline and do not represent a volume-averaged LyC escape fraction. With our results from \autoref{sec:results}, we can work to answer the three questions posed in \autoref{sec:intro}.


In \autoref{fig:fescvebmv_figure}, we observe significant spatial variations in the \ion{Mg}{II} escape fractions in J0919. The curve in \autoref{fig:fescvebmv_figure} represents the theoretical relationship between the \ion{Mg}{II} escape fraction and E(B-V) when only accounting for dust. Many of the points are suggestive of being described as being only related to dust attenuation, however the curve does not describe the full variation. This suggests that the \ion{Mg}{II} escape fraction depends on the dust attenuation, either through direct loss of photons due to increased dust absorption from increased dust column densities or increased dust absorption due to an increased photon path length as the \ion{Mg}{}$^+$ column density increases and the \ion{Mg}{II} resonantly scatters the photons. There are only four spatially distinct points here and a larger sample is needed to explore the full variation.

Since the dust attenuates LyC photons more than \ion{Mg}{II} photons (see \autoref{sub:IDE}) and is also correlated with the \ion{Mg}{II}/[\ion{O}{III}] ratio, this means that the LyC escape fraction varies spatially even more dramatically than the \ion{Mg}{II} escape fraction. For example, from the values in \autoref{tab:escapefractions}, the Western and Southern apertures differ by a factor of 2 in their \ion{Mg}{II} escape fraction and their LyC escape fractions differ by an order of magnitude. More concretely, the LyC escape fraction in the Southern aperture is 2\%, while it is approximately 72\% in the Western aperture.  Intriguingly, \ion{Mg}{II} and dust observations also imply that if we could measure the LyC escape fraction from a different line of sight within J0919, we would likely infer significantly different LyC escape values. This suggests that LyC escape fractions are highly sightline dependent and that sightline-to-sightline variations could introduce significant scatter to both indirect and direct estimates of LyC escape, as observed in simulations (e.g. \citealt{Trebitsch_2017,Rosdahl_2018,Mauerhofer_2021}).


Indirect tracers of LyC escape fractions, such as [\ion{O}{III}]/[\ion{O}{II}], may be less useful due to the scatter introduced by single sightline observations. To quantify this scatter we take the average of the \ion{Mg}{II} and LyC escape fractions from all the spatially distinct regions (North, South, West, and East; see \autoref{tab:escapefractions}). We estimate that single sightline observations of LyC escape can vary from one to ten times the average value. In stark contrast to this, there is only a factor of 1.3 difference in the [\ion{O}{III}]/[\ion{O}{II}] values. This fact implies that at fairly constant [\ion{O}{III}]/[\ion{O}{II}] values, the line of sight \ion{Mg}{II} escape, and by extension LyC escape, can take on an order of magnitude different values. Furthermore, our \ion{Mg}{II} observations suggest that the scatter in these relations are largely driven by spatial variations of dust and neutral gas. This order of magnitude scatter is fairly similar to the scatter that is observed in the indirect LyC trends of \citealt{Flury_2022b}. Recovering any trends from this amount of scatter would require very large samples to average over the spatial variations.  

\subsection{Single sightlines are unlikely to represent volume averages}
\label{sub:single sightlines}
Using the Integrated aperture to estimate the LyC escape can reduce the scatter in trends because it would average over many individual escape routes of ionizing photons (see \autoref{tab:escapefractions}). The spatially integrated value has the same value as the average of all of the single sightlines. With this being the case for this galaxy, one can estimate the average escape fraction from the mean of the observations. While this may be the case, we have to keep in mind that the Integrated aperture does not fully represent the "volume-averaged" escape fraction because we do not see the backside of the galaxy  (\autoref{tab:escapefractions}). 

The Central aperture, the brightest and one of the most transparent regions, represents the region that is commonly probed with single sightline observations. As mentioned above, this region is where the LyC measurement from \citet{Izotov_2021} was made. The question is then, how well does our brightest region represent the total escape fraction of the galaxy? The Central aperture has $\sim1.3$ times less \ion{Mg}{II} escape fraction and $\sim2$ times less LyC escape fraction when compared to the Integrated aperture. The values for the Central aperture do not agree with the average escape fraction of all the spatially distinct regions. For J0919, a single sightline observation, especially one of a bright commonly probed region, does not result in an accurate prediction of the average escape fraction of a galaxy.

Given how different some apertures are (e.g. the Southern and Western apertures), single sightline observations can introduce significant scatter, which makes it crucial to constrain the spatial variation of the escape fraction. For instance, if the Hubble Space Telescope had observed J0919 through a different region than the center, it could have estimated an escape fraction near 2\% (Southern aperture), which is no longer above the theorized threshold of 5-20\% to be a cosmically relevant source of reionization \citep{Ouchi_2009, Robertson_2013, Robertson_2015,Finkelstein_2019, Naidu_2020}. Thus, even if a single sightline implies large LyC escape fractions, the entire galaxy may not be emitting sufficient ionizing photons to reionize the Universe.


While a single sightline observation may not produce escape fractions that are representative of the entire galaxy, we can produce estimates of the amount of reionization-powering photons that escape J0919 per second to test the single sightline observation. To produce this estimate, referred to as Q\textsubscript{HI}, we use F(H$\beta$), Fesc\textsubscript{LyC}, and a constant, c\textsubscript{l}, along with eq. 1 from \citealt{Schaerer_2001} (see \autoref{tab:escapefractions}). H$\beta$ can trace the production of ionizing photons because in regions where an electron is ionized off its proton, the electron will recombine with another free proton to emit a Balmer emission line such as H$\beta$. From only the flux values in \autoref{tab:all_fluxes_table}, one might expect that the Central or Southern region would dominate in the emission of ionizing photons given that they have $\approx1.2-2$ times more F(H$\beta$) than the Western region. However, the Western region emits a total of seven times more ionizing photons than the Central region. Even though the Southern and Central regions produce more ionizing photons, the high escape fraction of the Western region means that more ionizing photons escape from the Western region. A caveat to this technique is that in regions that are very optically thin to ionizing photons, like the Western region, the ionizing photons will escape and not ionize the hydrogen, thus reducing the amount of F(H$\beta$). This means that our estimate of the total ionizing emissivity of the Western region is likely a lower limit, and the true number of ionizing photons that escape this region could be higher. There is a scatter of an order of magnitude around the average value.

From our estimates of the escape fractions and the photon counts, we can return with answers to the three questions posed  in \autoref{sec:intro}: (a) Single sightline observations do not necessarily provide the escape fraction of an entire galaxy, but on scales large enough to cover the spatial extent of the galaxy, like the Integrated aperture, the single sightline can get close to the average of the galaxy. (b) A single LyC detection does not represent a "volume-averaged" escape fraction and a single sightline can produce escape fractions that are far from the "volume-average". (c) Large spatial variations in LyC escape may lead to scatter in indirect estimators of LyC escape. These results rely on one galaxy and similar studies are needed to better investigate the sightline effect.  

\section{Summary and conclusions}
In this paper we presented LRS2 spatially resolved spectroscopic observations of \ion{Mg}{II}$\lambda2796$, \ion{Mg}{II}$\lambda2803$, [\ion{O}{II}]$\lambda\lambda3727,3729$, [\ion{Ne}{III}]$\lambda3869$, H$\gamma$, [\ion{O}{III}]$\lambda4363$, H$\beta$, [\ion{O}{III}]$\lambda4959$, [\ion{O}{III}]$\lambda5007$, and H$\alpha$ from the previously confirmed z$\approx$0.4 LyC emitting galaxy, J0919+4906 (\autoref{fig:fitexamples_figure}). J0919+4906 has an ionizing photon escape fraction of 16\% \citep{Izotov_2021}.

In order to test the spatial variation of \ion{Mg}{II} emission, dust attenuation, and nebular ionization, we separated our data into four spatially distinct apertures and one large aperture to contain all of the signal from our galaxy (\autoref{sub:aperture}). We include a central aperture and a COS analogue aperture to capture the brightest region of the galaxy and to compare our methods to the literature. This central region is where the LyC detection was made in \citet{Izotov_2021}. The spatially distinct apertures were separated by more than the convolved seeing of the observations (\autoref{fig:aperture_figure}).


We observe spatial variations in \ion{Mg}{II} emission, dust attenuation, and nebular ionization (Figures \ref{fig:overlay_figure}, \ref{fig:dustvmgii_figure}, and \ref{fig:o32vmgii_figure}). More specifically we find: regions with less dust attenuation are more ionized (\autoref{fig:o32vdust_figure}); regions with more \ion{Mg}{II} flux relative to [\ion{O}{III}]$\lambda5007$ are more highly ionized and have less dust attenuation (Figure \ref{fig:dustvmgii_figure}, \ref{fig:o32vmgii_figure}).

From our observed values for ionization and \ion{Mg}{II} emission taken relative to [\ion{O}{III}]$\lambda5007$, we use photoionization models to calculate the escape fraction of \ion{Mg}{II} (\autoref{sub:IDE}). We find that there is large spatial variation in the \ion{Mg}{II} escape fraction (\autoref{tab:escapefractions}). From these variations we find that regions with low dust attenuation and high ionization will have a larger fraction of the intrinsic \ion{Mg}{II} emission escape in the galaxy (\autoref{fig:fescvebmv_figure}). In \autoref{sub:aperture variations} we dust correct the \ion{Mg}{II} escape fractions to estimate the LyC escape fractions. The regions with low \ion{Mg}{II} escape, due to the strong correlation between \ion{Mg}{II} and dust, have even lower LyC escape fractions (\autoref{tab:escapefractions}).

We find that the Integrated aperture, which contains all of the signal from the galaxy, represents the average \ion{Mg}{II} and LyC escape fractions of the spatially distinct apertures (\autoref{tab:escapefractions}). In contrast, the Central aperture is the brightest region and thus the region where the LyC was observed. We were able to recover this observed LyC escape fraction by using the dust corrected \ion{Mg}{II} escape fraction from the COS analogue aperture. The Central/COS analogue aperture does not represent the average escape fraction. This places the average of the galaxy within the typically-quoted value needed to reionize the universe even though the Southern aperture value is below this limit. The variability in the values of each sightline could introduce scatter in the indirect estimators of LyC escape. With this information, we determine that single sightline observations may not accurately reflect the volume averaged LyC escape fraction. To resolve this issue, an extremely large sample size must average over many different sightline combinations. Combining a large sample size with observations on the scale of our Integrated aperture should reduce scatter and indirectly offer volume averaged escape fractions. It should be noted that the amount of scatter in spatially resolved \ion{Mg}{II} properties found within a galaxy has only been studied in a handful of galaxies (J0919 and J1503; \citealt{Chisholm_2020}) and will need to be validated on a larger sample.

The recently launched \textit{James Webb Space Telescope} (JWST) will offer observations of \ion{Mg}{II}, which has been redshifted into the IR for high-redshift galaxies. The size of these high-redshift galaxies on the sky means that they will fall within one micro-shutter assembly (MSA) and be probed similarly to our Central aperture. As this value may be far from the average for the galaxy, the scatter demonstrated in this paper due to single sightlines should be taken into account.

\section*{Acknowledgements}
We would like to thank the referee (Alaina Henry) for all of their insightful comments which greatly increased the quality of this paper.

We would like to acknowledge that the HET is built on Indigenous land. Moreover, we would like to acknowledge and pay our respects to the Carrizo \& Comecrudo, Coahuiltecan, Caddo, Tonkawa, Comanche, Lipan Apache, Alabama-Coushatta, Kickapoo, Tigua Pueblo, and all the American Indian and Indigenous Peoples and communities who have been or have become a part of these lands and territories in Texas, here on Turtle Island.

The Hobby-Eberly Telescope (HET) is a joint project of the University of Texas at Austin, the Pennsylvania State University, Ludwig-Maximilians-Universität München, and Georg-August-Universität Göttingen. The HET is named in honor of its principal benefactors, William P. Hobby and Robert E. Eberly.

The Low Resolution Spectrograph 2 (LRS2) was developed and funded by the University of Texas at Austin McDonald Observatory and Department of Astronomy and by Pennsylvania State University. We thank the Leibniz-Institut für Astrophysik Potsdam (AIP) and the Institut für Astrophysik Göttingen (IAG) for their contributions to the construction of the integral field units.

\section*{Data Availability}
The data underlying this article will be shared on request to the corresponding author.



\bibliographystyle{mnras}
\bibliography{refbibliography} 


\bsp	
\label{lastpage}
\end{document}